\documentclass[prd,twocolumn,superscriptaddress,nofootinbib
  ,noshowpacs,preprintnumbers
  ]{revtex4-1}
\usepackage[utf8]{inputenc}
\usepackage{lmodern}
\usepackage{amsmath,amssymb,mhchem}
\usepackage{graphicx}
\usepackage{url}
\usepackage{color}
\usepackage{enumitem}
\usepackage[dvipsnames]{xcolor}
\usepackage[colorlinks=true,breaklinks=true]{hyperref}
\hypersetup{allcolors=[rgb]{0.0 0.0 0.6},linkcolor=[rgb]{0.75 0.05 0.05}}
\usepackage{bm}
\usepackage{epsfig}
\usepackage{amsmath}
\usepackage{amssymb}
\usepackage{slashed}
\usepackage{color}
\usepackage{accents}
\usepackage[dvipsnames]{xcolor}
\usepackage[colorlinks=true,breaklinks=true]{hyperref}
\hypersetup{allcolors=[rgb]{0.0 0.0 0.6},linkcolor=[rgb]{0.75 0.05 0.05}}

\renewcommand\({\left(}
\renewcommand\){\right)}
\renewcommand\[{\left[}
\renewcommand\]{\right]}

\newcommand{\exclude}[1]{}

\newcommand\barparen{\scalebox{.3}{(}\raisebox{-1.2pt}[0pt][0pt]{$\mkern-1.3mu\scriptstyle-\mkern-1.3mu$}\scalebox{.3}{)}}

\begin{document}

\preprint{MPP-2019-67}

\title{Distinguishing Dirac and Majorana neutrinos by their gravi-majoron decays}

\author{Lena Funcke}
\affiliation{Perimeter Institute for Theoretical  Physics,
31 Caroline Street North, Waterloo, Ontario, N2L 2Y5, Canada}

\author{Georg Raffelt}
\affiliation{Max-Planck-Institut f\"ur Physik (Werner-Heisenberg-Institut),
  F\"ohringer Ring 6, 80805 M\"unchen, Germany}

\author{Edoardo Vitagliano}
\affiliation{Max-Planck-Institut f\"ur Physik (Werner-Heisenberg-Institut),
  F\"ohringer Ring 6, 80805 M\"unchen, Germany}

\date{\today}

\begin{abstract}
Neutrinos may acquire small Dirac or Majorana masses by new low-energy physics in terms of the chiral gravitational anomaly, as proposed by Dvali and Funcke (2016). This model predicts fast neutrino decays, $\nu_i\to\nu_j+\phi$ and $\nu_i\to\overline{\nu}_j+\phi$, where the gravi-majorons $\phi$ are pseudoscalar Nambu-Goldstone bosons. The final-state neutrino and antineutrino distributions differ depending on the Dirac or Majorana mass of the initial state. This opens a channel for distinguishing these cases, for example in the spectrum of high-energy astrophysical neutrinos. In particular, we put bounds on the neutrino lifetimes in the Majorana case, ${\tau_2}/{m_2}> 1.1\times 10^{-3}(6.7\times 10^{-4})$~s/eV and ${\tau_3}/{m_3}> 2.2\times 10^{-5}(1.3\times 10^{-4})$~s/eV at 90\% CL for hierarchical (degenerate) masses, using data from experiments searching for antineutrino appearance from the Sun.
\end{abstract}

\maketitle

\section{Introduction}

A completely new approach to explain small Dirac or Majorana neutrino
masses \cite{Dvali2016a} relies on new physics at the low-energy
frontier of particle physics instead of high-energy extensions of the
Standard Model. The key idea is that a hypothetical topological vacuum
susceptibility of gravity induces fermion condensation \citep{Dvali:2005an,Dvali:2013cpa}, which can give rise to effective
fermion masses. If this effect is realized in nature, it is most
important for the lightest fermions and could be the unique
origin of neutrino masses.

Phenomenologically, neutrino condensation would be accompanied by the
appearance of pseudoscalar Nambu-Goldstone bosons $\phi$,
similar to light mesons in QCD, that can be interpreted as
neutrino-antineutrino bound states~\citep{Dvali:2013cpa,Dvali2016a} and that we call
gravi-majorons.\footnote{This suggestive terminology is a bit of a
  misnomer because majorons \cite{Chikashige1981} were originally invoked to explain
  Majorana masses, whereas it is a key point of our discussion that
  $\phi$ bosons appear for both the Dirac and Majorana option.}  An important difference to conventional Nambu-Goldstone bosons is that the $\nu\nu\phi$ vertex, due to its low-energy origin, ``melts'' for high-energy off-shell
situations so that constraints based on scattering processes,
e.g., majoron production in supernovae, typically do not apply \cite{Dvali2016a,Dvali2016b}. On the
other hand, decays of the type $\nu_i\to \overset{\barparen}{\nu}_j+\phi$ proceed in the
usual way.

While this scenario is exotic, it may become
empirically motivated in the near future. One predicts a neutrinoless
Universe after photon decoupling, or at least all neutrinos in the lowest
mass state, due to decays and annihilations into gravi-majorons \citep{Dvali2016a}. If
near-future cosmological observations \cite{Brinckmann:2018owf} fail to detect
a hot dark matter component
on the minimal level expected from oscillation experiments, we may be
forced to contemplate the absence of the usual cosmic neutrino
background.  Similar questions arise if the KATRIN experiment \cite{Drexlin2013,Aker:2019uuj} detects
a neutrino mass in conflict with cosmological limits.

The gravitational mass model works for both Dirac and
Majorana neutrinos. Therefore, one important question remains how we
can experimentally distinguish between these two possibilities.  In
high-energy models, neutrinoless double-beta
($0\nu\beta\beta$) decay \citep{Dolinski:2019nrj} is the most promising approach and one that
remains viable in our scenario. Moreover, in the current paper we predict that the low-energy gravitational mass model offers an additional opportunity through fast $\nu_i\to \overset{\barparen}{\nu}_j+\phi$
decays.

It is not
new that details, e.g., of radiative decays $\nu_i\to \overset{\barparen}{\nu_j}+\gamma$ depend on the Dirac vs.\ Majorana nature
\cite{Balantekin:2018azf,Balantekin:2018ukw}. The $\gamma$ spectrum
in relativistic decays depends on this property, inherited from the
angular $\gamma$ distribution relative to the spin of the mother
neutrino in its rest frame. However, radiative decays of light neutrinos are usually
too slow to be of any practical interest.

We argue that the $\nu_i\to \overset{\barparen}{\nu}_j+\phi$ decays in the gravitational mass model are in principle  fast enough to distinguish between the Majorana and Dirac cases by using the flux and spectrum of the daughter neutrinos. A Dirac
neutrino $\nu_i$ decays into a neutrino, either active $\nu_j$ or sterile $N_j$, whereas a
Majorana neutrino always decays into an active state, which however in
a detector appears as either a neutrino $\nu_j$ or an antineutrino $\overline{\nu}_j$. This is possible because a Majorana neutrino does not have a defined lepton number, and what we call antineutrino is simply a state with right-handed helicity. 

A detector that can distinguish neutrinos from
antineutrinos could identify the Dirac or Majorana nature by looking at
neutrino vs.\ antineutrino appearance, assuming an asymmetry at the
source.  Moreover, given the source spectrum, the energy spectrum
depends on the nature of the mass term independently of an asymmetry at the source.  If the mass spectrum is
degenerate, Majorana and Dirac particles could be distinguished because
spin-flip is not suppressed compared to spin conservation. As a result,
Majorana neutrinos would decay to antineutrinos, whereas Dirac
neutrinos would decay to sterile states. While these methods are not yet experimentally feasible, with the current data we can put strong bounds on the Majorana case using null results from experiments searching for antineutrino appearance from the Sun.

The paper is structured as follows. In Sec.~\ref{sec:previousmodels} we explain the key phenomenological differences between conventional majoron-like models and the gravi-majoron scenario. In Sec.~\ref{sec:decayrate} we compute the neutrino decay rate and in Sec.~\ref{sec:decayconstraints} we comment on their impact on solar, IceCube, and supernova neutrinos. We summarize our findings in Sec.~\ref{sec:conclusions}.

\section{Differences to traditional majoron-like models}
\label{sec:previousmodels}

\subsection{Dirac or Majorana neutrino nature}\label{sec:nunature}

In contrast to the original majoron model \cite{Chikashige1981}, the gravitational neutrino mass mechanism works for both Dirac and Majorana neutrinos. Therefore we can in principle distinguish three possible scenarios:

\begin{enumerate}[label={(\arabic*)}]
\item \textit{Pure left-handed Majorana case:} If neutrinos are identical to their antiparticles and no right-handed states exist, neutrinos are massless in the Standard Model. The neutrino condensate then generates small left-handed Majorana masses.
\item \textit{Pure Dirac case:} If neutrinos are distinct from their antiparticles, right-handed states exist and the condensate generates small effective Dirac masses.
\item \textit{Mixed case:} If neutrinos are identical to their antiparticles and right-handed states exist, the condensate generates Dirac masses as well as left- and right-handed Majorana masses. In this case, the active and sterile Dirac states have masses~$m_\nu\sim$~meV-- eV and are substantially mixed.
\end{enumerate}
For simplicity, our computations in Sec.~\ref{sec:decayrate} will focus on the minimal cases (1) and (2). In the former case, the effective left-handed Majorana mass terms violate isospin by one unit but are allowed after electroweak symmetry breaking. Here, we note that only the condensate but not gravity is assumed to violate isospin. In the latter case, the Yukawa couplings to the SM Higgs doublet are assumed to vanish based on chiral symmetry protection, so that the gravitational mechanism is the only origin of the observed neutrino masses. Note, however, that even in the presence of nonzero Yukawa couplings or other hard neutrino mass origins (e.g., the various seesaw mechanisms~\cite{King:2003jb}), neutrino condensation would still take place as long as $m_{\rm \nu, hard}<\Lambda_G$. Here, $\Lambda_G$ is the scale of the topological vacuum susceptibility of gravity, which is constrained by model-independent phenomenological requirements to lie within the range of $\Lambda_G\sim$~meV-- eV~\cite{Dvali2016a,Dvali2016b,Lorenz:2018fzb}. Since the susceptibility induces fermion condensation, $\Lambda_G$ also sets the neutrino condensation scale, $\langle\bar{\nu}_i\nu_j\rangle\sim\Lambda_G^3$, and thus the order of the absolute neutrino mass scale (see Ref.~\cite{Dvali2016a} for a discussion of the hierarchy).

Concerning the gravi-majorons, the pure Dirac case~(2) gives rise to 14+1 (pseudo) Nambu-Goldstone bosons  $\phi\equiv\{\phi_{k},\eta_\nu\}$ from spontaneously breaking the neutrino flavor symmetry $U(3)_V\times U(3)_A\rightarrow U(1)_V^3$. Here, the $\eta_\nu$ boson comes from the anomalous $U(1)_A$ part~\citep{Dvali:2013cpa} and the 14 $\phi_k$ bosons come from the $SU(3)_V\times SU(3)_A$ part~\cite{Dvali2016a}. The $\eta_\nu$ gets a mass through the chiral gravitational anomaly, $m_{\eta_\nu}\sim \Lambda_G\sim T_G$, while the $\phi_k$ are massless if neutrinos have only effective (gravitational) masses and no hard masses.\footnote{This discussion neglects tiny corrections for some $\phi_k$ masses due to weak $SU(2)_W$ effects, $m_{\phi_k}\sim 10^{-33}~{\rm eV} (\Lambda_G/{\rm 0.1~eV})^3$.} The analogous left-handed Majorana case~(1) would yield fewer bosons due to breaking a smaller initial flavor symmetry.

The off-diagonal couplings between gravi-majorons and neutrinos are responsible for neutrino decay. In typical majoron models such couplings are heavily suppressed by $(m_{\rm \nu,Dirac} / m_{\rm \nu,Majorana})^4 \ll 1$ because the majorons only couple to the heavy right-handed neutrino states~\cite{Schechter1982}. In our case, there is no such suppression and we can treat the off-diagonal couplings as free parameters, which are proportional to the unknown entries of the neutrino-mass diagonalization matrix.

To briefly comment on the mixed case (3), this active-sterile neutrino scenario could be motivated by short-baseline anomalies~\cite{Gariazzo2015}. While light sterile neutrinos generally conflict with cosmological constraints on neutrino masses and the effective number of neutrino species \cite{Adhikari2016}, these tensions disappear in the gravitational mass model because (i)~the mass bounds are weakened \cite{Dvali2016a,Lorenz:2018fzb} and (ii)~the active relic neutrino background is massless before photon decoupling and therefore uncoupled to the sterile states (see Appendix \ref{app:softening} for more details).

We note that light sterile neutrinos would strongly distort the parameter space for $0\nu\beta\beta$ decay \citep{Girardi2013} and could even make the decay vanish completely \cite{deGouvea2005}. In contrast, gravi-majorons do not affect the $0\nu\beta\beta$ decay because their emission is strongly suppressed (see Sec.~\ref{sec:astrobounds}). Finally, there was a debate\footnote{We thank Nima Arkani-Hamed for raising this question and Gia Dvali and Misha Shifman for further discussions. We also note that this question has not yet been unambiguously answered.} whether the $0\nu\beta\beta$ decay rate might be altered because the gravitational neutrino masses are not hard masses generated at high-energy scales but effective ones generated at the low-energy condensation scale $\Lambda_G$. However, since there is no momentum flow through the masses generated by the condensate, we expect them to be indistinguishable from hard neutrino masses in the $0\nu\beta\beta$ process.

\subsection{Late neutrinoless Universe}\label{sec:cosmobounds}

Interactions between neutrinos and Nambu-Goldstone bosons are strongly constrained by cosmological data. For example, almost the entire parameter space of the ``neutrinoless Universe'' model \cite{Beacom2004}, which evades the cosmological bounds $\sum_i m_{\nu_i} \lesssim 0.2~{\rm eV}$ \cite{Cuesta2015}, was ruled out by early-Universe neutrino free-streaming constraints \cite{Hannestad2004,Follin:2015hya,Lancaster:2017ksf,Oldengott:2017fhy,Kreisch:2019yzn} and precision measurements of the primordial radiation density \cite{Aghanim:2018eyx}.
In contrast, gravi-majorons are not ruled out because they only arise in the late Universe after photon decoupling. We will discuss the details of this high-temperature suppression of the new gravitational effects in Appendix \ref{app:softening}.

When the Universe cools down to the low phase-transition temperature~$T_G\lesssim T_{\rm decoupling}\sim 0.3~{\rm eV}$, the neutrino condensate forms and the emerging gravi-majorons $\phi\equiv\{\phi_{k},\eta_\nu\}$ \citep{Dvali:2013cpa,Dvali2016a} start to interact with neutrinos through the same nonperturbative gravitational vertex that is responsible for the neutrino masses,
\begin{align}\label{eq:nudecay}
\hspace*{-0.1cm} \mathcal{L}_{\rm int} =  &\sum_{k=1}^{8}  \partial_\mu \phi_k \sum_{i,j=1}^{3} \frac{g_{ij,k}}{m_i+m_j} \overline{\nu}_i \gamma^\mu\gamma_5 \nu_j \nonumber
\\
&+\sum_{k=9}^{14}  \partial_\mu \phi_{{k}} \sum_{i\neq j=1}^{3} \frac{{g}_{ij,k}}{m_i-m_j} \overline{\nu}_i \gamma^\mu \nu_j \nonumber
\\
&+ i\eta_\nu \sum_{i,j=1}^{3} g_{ij,15} \overline{\nu}_i \gamma_5 \nu_j+ \rm{h.c.} 
\end{align}
Here, the dimensionless couplings $g_{ij,k}$ are normalized in such a way that we recover the pseudoscalar-coupling Lagrangian in Eq.\ \eqref{eq:nudecay2}. The couplings correspond to the generators of the broken flavor symmetry, i.e., the Gell-Mann matrices plus the identity. The $U(3)_A$ generators yield diagonal and off-diagonal axial vector-like couplings proportional to $m_i+m_j$ (affecting 8+1 of the 14+1 bosons), while the $U(3)_V$ generators yield off-diagonal vector-like couplings proportional to $m_i-m_j$ (affecting the other 6 bosons), similar to familon models (see, e.g., Ref.~\cite{Feng1998}). Note that these couplings do not yield long-range forces in macroscopic systems, since the pseudoscalar coupling is spin-dependent and the scalar coupling is flavor-changing. In the following, we neglect the sum over $k$ and conservatively assume that the couplings are to a single Nambu Goldstone boson.

After the cosmological phase transition, the previously massless neutrinos become massive, quickly decay into the lowest mass eigenstate, and annihilate into gravi-majorons through the process $\nu + \overline{\nu}\rightarrow \phi + \phi$ \cite{Dvali2016a}. Thus, the early-Universe massless neutrino ``radiation'' converts into massless gravi-majoron radiation after photon decoupling, since all massive $\phi$ bosons decay into massless ones. This almost complete annihilation could only be evaded in the hypothetical presence of neutrino asymmetries in the Dirac case, weakening the cosmological neutrino mass bounds to $\sum m_\nu < 4.8$~eV at 95\% CL~\cite{Lorenz:2018fzb}.

To conclude, unless there are large primordial neutrino asymmetries, the gravitational neutrino mass model~\cite{Dvali2016a} predicts a late neutrinoless Universe and could be falsified by a cosmological neutrino mass detection, e.g., by the upcoming DESI or Euclid surveys \cite{Brinckmann:2018owf}. The KATRIN beta-decay experiment \cite{Drexlin2013,Aker:2019uuj} could provide a hint toward our predicted neutrinoless Universe if it detects an unexpectedly large neutrino mass scale. We stress that KATRIN's measurement of the electron energy spectrum would be unaffected by the gravi-majorons (unlike the bosons considered in Ref.~\cite{Arcadi:2018xdd}) because their interactions take place on much longer timescales than beta decay. The neutrinoless Universe scenario would also make it impossible for the recently proposed PTOLEMY experiment~\cite{Betts:2013uya} to detect the relic neutrino background. Only in the case of substantial neutrino asymmetries, PTOLEMY could detect the relic neutrinos with an enhanced (suppressed) detection rate for normal (inverted) neutrino mass ordering, since the lightest mass eigenstate contains a large (small) fraction of the electron-neutrino flavor eigenstate.\\

\subsection{Stellar and laboratory bounds}\label{sec:astrobounds}

One might naively expect that gravi-majorons are ruled out by astrophysical observations because their couplings are severely constrained by stellar processes. As explained in Ref.~\cite{Dvali2016b}, the least suppressed two-photon coupling of the gravi-majorons is $g_{\phi\gamma}\sim(\Lambda_G/m_e)^3 /\Lambda_G \sim 10^{-10}~{\rm GeV^{-1}}$ for $\Lambda_G\sim 0.1$~eV, which at first sight is already ruled out by constraints from solar axion experiments, $g_{\phi\gamma}\lesssim 0.88\times 10^{-10}~{\rm GeV^{-1}}$ at 95\% CL for $m_\phi\lesssim 0.02$~eV \cite{Anastassopoulos:2017ftl}. However, the gravi-majoron production in astrophysical environments with $E\gg m_\nu$ is additionally suppressed by $(\Lambda_G/E)^n$ due to the high-energy softening of the gravitational vertex, where $n$ is an unknown power-law exponent that has to be fixed by phenomenological requirements (as discussed in \cite{Dvali2016a,Dvali2016b} and App.\ \ref{app:softening}). Thus, the common relation between axion-like predictions for laboratory and solar axion experiments does not apply \cite{Dvali2016b}. By similar arguments one also evades other astrophysical bounds on the gravitational $\nu\overline{\nu}\gamma$, $\nu\overline{\nu}\phi$, $\nu\overline{\nu}\nu\overline{\nu}$, $e\overline{e}\phi$, and $e\overline{e}\nu\overline{\nu}$ couplings, as we comment on in Appendix \ref{app:softening}.

Regarding laboratory experiments, the most important bounds on the emission of conventional majoron-like bosons comes from $0\nu\beta\beta$ decay \cite{Arnold2016} and leptonic decays of mesons \cite{Pasquini2015}. Following the argumentation above, these bounds do not apply to gravi-majorons because they and their couplings to the virtual intermediate neutrinos dissolve in these high-energy processes. However, short-distance fifth-force experiments can put important bounds on our new gravitational interactions, as discussed in \citep{Dvali2016a,Dvali2016b} and Appendix \ref{app:softening}.

We finally stress that the gravitational vertex suppression is due to the large four-momentum transfer in the processes under consideration, which is why the suppression does not apply to neutrino oscillations or to the neutrino decays considered in Sec.~\ref{sec:decayrate}.

\section{Neutrino Decay Rate}
\label{sec:decayrate}

As predicted by the gravitational neutrino mass model \citep{Dvali2016a}, the presence of gravi-majorons implies neutrino decay through the vertex of Eq.~\eqref{eq:nudecay}. The crucial new point of this paper is that the Dirac and Majorana neutrino cases yield different decay channels, so that the composition of the daughter neutrinos depends on the type of masses generated through the gravitational mechanism.

\subsection{Pseudoscalar decay channels}

For the two neutrino decay channels, $\nu_i\rightarrow\nu_j+\phi$ and $\nu_i\rightarrow\overline{\nu}_j+\phi$ with $m_i>m_j$, the axial derivative coupling in Eq.~\eqref{eq:nudecay} is equivalent to the pseudoscalar coupling \cite{Hannestad2005}
\begin{align}\label{eq:nudecay2}
 \mathcal{L}_{\rm int} = i \phi \sum_{ij} g_{ij} \overline{\nu}_i \gamma_5 \nu_j+\rm h.c. \qquad
\end{align}
as we have checked explicitly in Appendix \ref{appendix}. Notice the Lagrangian in Refs.~\cite{Kim:1990km, Lindner:2001fx, Moss:2017pur} includes a factor of $1/2$.

The differential rate $\Gamma$ for the decay $\nu_i\to \overset{\barparen}{\nu}_j + \phi$ in a generic reference frame is
\begin{align}
d\Gamma=\frac{1}{2 E_i}(2\pi)^4\delta^{(4)}(p_i-p_j-k)|\mathcal{M}|^2\frac{d^3 \bold{p}_j}{(2\pi)^3 2E_j}\frac{d^3 \bold{k}}{(2\pi)^3 2\omega}\ ,
\end{align}
where $p_i$, $p_j$ and $k$ are the four-momenta of $\nu_i$, $\nu_j$ and $\phi$, respectively. The squared amplitudes for both the helicity-preserving and the helicity-changing processes are given in Appendix \ref{appendix}. After integration over the gravi-majoron momenta and the direction of the neutrino, the differential rate becomes
\begin{align}
\frac{d\Gamma}{d E_j}=\frac{1}{16\pi E_i |\mathbf{p}_i|}|\mathcal{M}|^2  \ .\end{align}
In the laboratory frame, where $E_i\gg m_i$, we find for Majorana neutrinos
\begin{subequations}
\begin{align}
\frac{d\Gamma^\mathrm{M}_{\nu_i\to\nu_j+ \phi}}{d E_j}&=\frac{ g_{ij}^2}{4\pi E_i |\mathbf{p}_i|}  m_i m_j\(A-2\) \label{nunumajo} &&\\
\frac{d\Gamma^\mathrm{M}_{\nu_i\to\overline{\nu}_j+ \phi}}{d E_j}&=\frac{ g_{ij}^2}{4\pi E_i |\mathbf{p}_i|} m_i m_j\(\frac{m_i^2+m_j^2}{m_i m_j}-A\) \ , \label{nuantinumajo}
  \end{align}
  \end{subequations}
where
\begin{align}
A\equiv \( \frac{m_i E_j}{m_j E_i}+\frac{m_j E_i}{m_i  E_j }     \)   \ .
\end{align}
Dirac neutrinos decay either into active (left-handed) neutrinos or into sterile (right-handed) neutrinos $N_j$. The differential decay rates equal $1/4$ times the rates for Majorana neutrinos,
\begin{subequations}
\begin{alignat}{2}
\frac{d\Gamma^\mathrm{D}_{\nu_i\to\nu_j+ \phi}}{d E_j}&=\frac{ 1}{4}  \frac{d\Gamma^\mathrm{M}_{\nu_i\to\nu_j+ \phi}}{d E_j} &&\\
\frac{d\Gamma^\mathrm{D}_{\nu_i\to N_j+ \phi}}{d E_j}&=\frac{ 1}{4} \frac{d\Gamma^\mathrm{M}_{\nu_i\to\overline{\nu}_j+ \phi}}{d E_j} \ . 
\end{alignat}
\end{subequations}
\begin{widetext}
The differential rates must be integrated over the allowed energy range for the daughter neutrino $\nu_j$,
\begin{align}
\frac{E_i}{2}\(1+\frac{1}{x_{ij}^2}\)- \frac{|\bold{p}_i|}{2}\(1-\frac{1}{x_{ij}^2}\) \le E_j\le  \frac{E_i}{2}\(1+\frac{1}{x_{ij}^2}\)+ \frac{|\bold{p}_i|}{2}\(1-\frac{1}{x_{ij}^2}\)
 \label{enrange}
\end{align}
where $x_{ij}\equiv~m_i/m_j>1$. In the lab frame with $E_i\gg m_i$, Eq.~\eqref{enrange} reduces to
\begin{align}
 \frac{E_i}{x_{ij}^2}\le E_j\le E_i \ .
\end{align}
The decay rates for Majorana neutrinos are then
\begin{subequations}
\begin{align}
\Gamma^\mathrm{M}_{\nu_i\to \nu_j+ \phi}&=\ \frac{g_{ij}^2 m_i m_j}{4\pi E_i}\(\frac{x_{ij}}{2}-2+\frac{2}{x_{ij}}\log x_{ij}+\frac{2}{x_{ij}^2}-\frac{1}{2x_{ij}^3}\) \ 
\\
\Gamma^\mathrm{M}_{\nu_i\to \overline{\nu}_j+ \phi}&=\ \frac{g_{ij}^2 m_i m_j}{4\pi E_i}\(\frac{x_{ij}}{2}-\frac{2}{x_{ij}}\log x_{ij}-\frac{1}{2x_{ij}^3}\) \ .
\end{align}
\end{subequations}
Our results agree with the literature \cite{Kim:1990km, Lindner:2001fx, Moss:2017pur} if one accounts for the different normalization of the couplings.

\clearpage
\pagebreak
\end{widetext}

There are two important limits for these expressions: the case of a large mass hierarchy $m_i\gg m_j$ and the degenerate case $m_i\simeq m_j$. The latter case remains allowed in the gravitational mass model because the cosmological bounds on neutrino masses are substantially weakened (Sec.~\ref{sec:cosmobounds}). In the limit of a large mass hierarchy, one finds
\begin{align}
\Gamma^\mathrm{M}_{\nu_i\to \nu_j+ \phi}=\Gamma^\mathrm{M}_{\nu_i\to \overline{\nu}_j+ \phi}= \frac{\Gamma^{\mathrm{M}}_{\mathrm{tot}}}{2}\ ,
\end{align}
where the total decay rate is
\begin{align}\label{eq:NDecay}
\Gamma^\mathrm{M}_\mathrm{tot}=\frac{g_{ij}^2 m_i}{4\pi} \frac{m_i}{E_i}\ .
\end{align}
Here, the first factor is the total decay rate in the rest frame of $\nu_i$ and $m_i/E_i$ is the Lorentz factor.
 
The total decay rate in Eq.~\eqref{eq:NDecay} yields the neutrino rest-frame lifetimes $\tau_i=1/\Gamma_i$ of
\begin{subequations}
\begin{align}
\frac{\tau_3}{m_3}&\simeq \frac{4\times 10^{-11}}{g_{32}^2+g_{31}^2}\frac{{\rm s}}{{\rm eV}}\ \\
\frac{\tau_2}{m_2}&\simeq \frac{1\times 10^{-9}}{g_{21}^2}\frac{{\rm s}}{{\rm eV}} \ .
\end{align}
\end{subequations}
For example, the smallest possible, normal-ordered neutrino masses of $m_1=0~{\rm meV}$, $m_2=9~{\rm meV}$, and $m_3=50~{\rm meV}$ \cite{Tanabashi:2018oca} give
\begin{subequations}
\begin{align}
\tau_3&\simeq \frac{2\times 10^{-12} \, \rm{s}}{g_{32}^2+g_{31}^2}\ \\
\tau_2&\simeq \frac{9\times 10^{-12} \, \rm{s}}{g_{21}^2}\ .
\end{align}
\end{subequations}
The degenerate limit, $m_i\simeq m_j$, gives
\begin{subequations}
\begin{align}
&\Gamma^\mathrm{M}_{\nu_i\to \overline{\nu}_j+ \phi}=2\Gamma^\mathrm{M}_{\nu_i\to \nu_j+ \phi}=\frac{g_{ij}^2 (m_i-m_j)^3}{3\pi m_i^2} \frac{m_i}{E_i} \ .
\end{align}
\label{eq:pseudoscalar_deg}
\end{subequations}

\subsection{Scalar decay channels}

The scalar decay channels arising from the derivative vector-like couplings in Eq.\ \eqref{eq:nudecay} are studied in analogous fashion. In this case, the derivative vector-like coupling can be computed using a scalar coupling,
\begin{align}
\mathcal{L}_{\rm int} = i \phi \sum_{i\neq j} g_{ij} \overline{\nu}_i \nu_j+\rm h.c.\,,
\end{align}
analogous to the pseudoscalar coupling in Eq.~\eqref{eq:nudecay2}.

The decay rates for Majorana neutrinos are 
\begin{widetext}
\begin{subequations}
\begin{align}
\Gamma^\mathrm{M}_{\nu_i\to \nu_j+ \phi}&=\ \frac{g_{ij}^2 m_i m_j}{4\pi E_i}\(\frac{x_{ij}}{2}+2+\frac{2}{x_{ij}}\log x_{ij}-\frac{2}{x_{ij}^2}-\frac{1}{2x_{ij}^3}\) \ 
\\
\Gamma^\mathrm{M}_{\nu_i\to \overline{\nu}_j+ \phi}&=\ \frac{g_{ij}^2 m_i m_j}{4\pi E_i}\(\frac{x_{ij}}{2}-\frac{2}{x_{ij}}\log x_{ij}-\frac{1}{2x_{ij}^3}\) \, .
\end{align}
\end{subequations}
\end{widetext}
Similar to the pseudoscalar case, the differential decay rates for Dirac neutrinos equal $1/4$ times the rates for Majorana neutrinos. In the limit of a large mass hierarchy, one finds again
\begin{align}
\Gamma^\mathrm{M}_{\nu_i\to \nu_j+ \phi}=\Gamma^\mathrm{M}_{\nu_i\to \overline{\nu}_j+ \phi}= \frac{\Gamma^{\mathrm{M}}_{\mathrm{tot}}}{2}\ 
\end{align}
where the total decay rate is
\begin{align}
\Gamma^\mathrm{M}_\mathrm{tot}=\frac{g_{ij}^2 m_i}{4\pi} \frac{m_i}{E_i}\ ;
\end{align}
this is identical to the decay rate found for the pseudoscalar decay channels. 
The degenerate limit, $m_i\simeq m_j$, gives instead
 \begin{align}
\Gamma^{\mathrm{M(D)}}(\nu_i\to {\nu}_j+ \phi)=\frac{g^2 }{\zeta^{\mathrm{M(D)}}\pi} \frac{(m_i^2-m_j^2)}{E_i} \
\end{align}
where $\zeta^{\mathrm{M}}=1$ and $\zeta^{\mathrm{D}}=4$, while the spin-flip process is strongly suppressed by a factor of $[(m_i^2-m_j^2)/m_i^2]^3$~\cite{Beacom:2002cb}. We observe that the scalar decay rate of Majorana (M) and Dirac (D) neutrinos in Eq.~(2.6) of Ref.~\cite{Beacom:2002cb} needs to be corrected. Our result corrects for an inconsistency between the coupling definition in their Lagrangian and their rate (if neutrinos are Majorana particles), as well as an incorrect expansion of their Eq.~(2.2).\footnote{We thank Nicole Bell for communications on this point.}

In the degenerate mass scenario, the main difference between the scalar and pseudoscalar decay channels is that helicity-flipping processes are strongly suppressed in the former case but not in the latter, see Eq.\ \eqref{eq:pseudoscalar_deg}. Since spin-flip processes are unsuppressed in the pseudoscalar case and their rate is different from the spin-conserving processes, we can identify several different observables to distinguish Majorana and Dirac neutrinos, as we will discuss in Sec.~\ref{sec:decayconstraints}. Given that the vector-like coupling is either equivalent to the axial vector-like coupling (in the hierarchical scenario) or cannot distinguish Majorana from Dirac neutrinos (in the degenerate scenario), we will focus for the rest of the paper on the axial vector-like couplings, i.e., the pseudoscalar decay channels.

\begin{figure}[hbt!]
 \vspace*{-1cm}

\includegraphics[width=0.85\columnwidth]{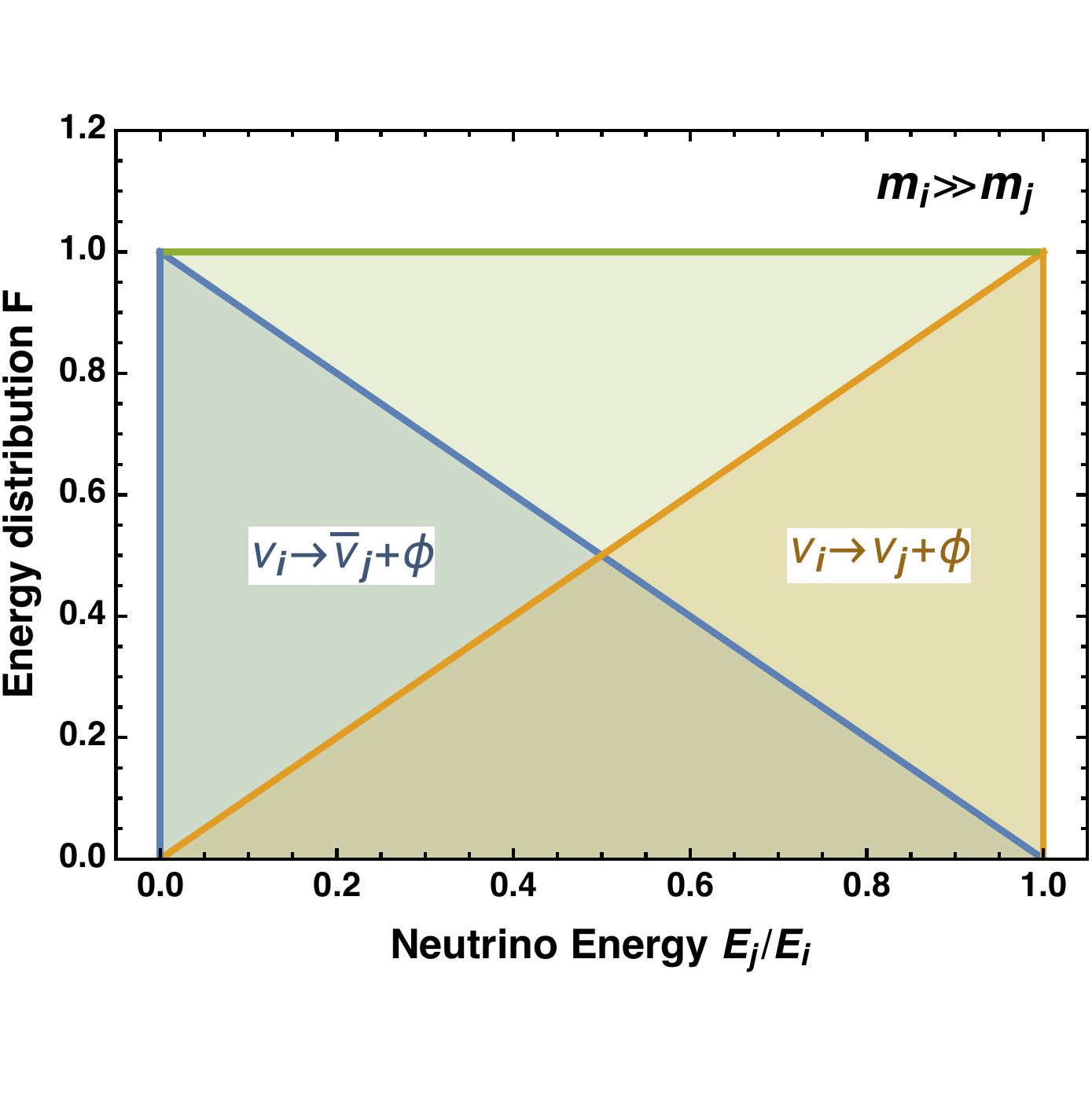}

 \vspace*{-1cm}

\includegraphics[width=0.85\columnwidth]{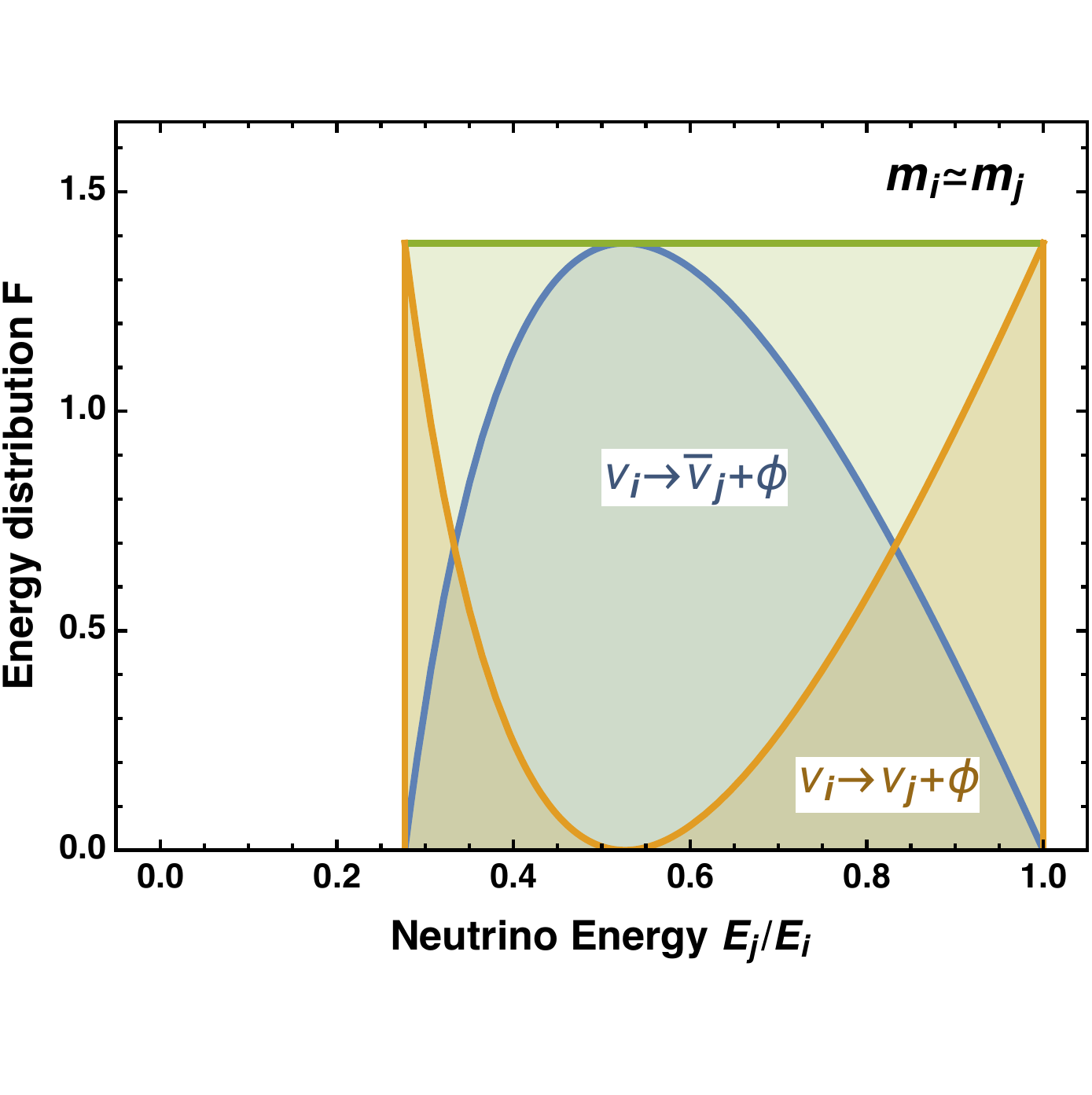}

 \vspace*{-1cm}
\caption{Normalized energy distributions $F(E_j/E_i)$ of the daughters ${\nu}_j$ and $\overline{\nu}_j$ produced in the pseudoscalar decay of Majorana neutrinos $\nu_i\rightarrow \nu_j + \phi$ (orange), $\nu_i\rightarrow \overline{\nu}_j + \phi$ (blue), and their sum (green), for hierarchical (upper panel) and degenerate (lower panel) neutrino masses. For the Dirac case, the blue curve instead refers to the decay to undetectable sterile states $N_j$.}
\label{fig:fig1}
\end{figure}

To show how the cases for Majorana and Dirac neutrinos are different, in Fig.~\ref{fig:fig1} we plot the energy distributions ${F(E_j/E_i)}$ of the daughter neutrinos produced in the pseudoscalar decay of the parent neutrinos $\nu_i$, which is proportional to the differential rate and normalized to the total decay rate. To describe the figure, let us begin with the decay of Majorana neutrinos with energy $E_i$. In the upper panel of Fig.~\ref{fig:fig1}, we show the hierarchical case, $m_{i}\gg m_{j}$. The spectrum of neutrinos and antineutrinos is a box given by the sum of two triangles, one corresponding to outgoing $\nu_j$ and one to outgoing $\overline{\nu}_j$. In the degenerate case $m_{i}\simeq m_{j}$ shown in the lower panel of Fig.~\ref{fig:fig1}, the two triangles are distorted, but the sum of the $\nu_j$ and $\overline{\nu}_j$ distributions is still a box, as the sum of Eqs.~\eqref{nunumajo} and~\eqref{nuantinumajo} is independent of energy. In the Dirac case, the spectrum of active daughter neutrinos is only the orange triangle because the complement to the box, the blue triangle, corresponds to undetectable sterile daughter neutrinos.

\section{Constraints and detection opportunities}\label{sec:decayconstraints}

One of the possibilities to distinguish Majorana from Dirac neutrinos is given by the appearance or disappearance of neutrinos or antineutrinos from a source whose flux is asymmetric between $\nu$ and $\overline{\nu}$. For example, the Sun emits only neutrinos \cite{Giunti:2007ry}, besides a tiny contribution of antineutrinos coming from heavy-element decay, which is much smaller than the geoneutrino background \cite{Vitagliano:2019yzm,1990ApJ...352..767M,Mantovani:2003yd}. The detection of antineutrinos could point toward the decay of Majorana neutrinos. Notice, however, that other explanations for such a detection would be possible, e.g., a spin precession due to a neutrino magnetic dipole moment \cite{Akhmedov:2002mf}. New liquid scintillator detectors will significantly increase the current best experimental limits~\cite{Li:2019snw}.

A second possibility relies on the observation of the differential energy spectrum from a source whose flux is known. In this case, there is no need for an asymmetry between the initial neutrino and antineutrino fluxes because the differential energy spectrum is differently affected by the decay depending on the nature of the neutrino mass.

Finally, the observation of a change in the flavor pattern in the neutrino flux from a known source could in principle distinguish Majorana and Dirac neutrinos if it is possible to distinguish neutrinos from antineutrinos in the detector. Otherwise, it is only possible to observe the decay, as the latter changes the neutrino composition in terms of mass eigenstates of the flux.

In the following, we will quantitatively examine the first possibility, which yields bounds on the Majorana case of our neutrino decay scenario. Moreover, we will qualitatively discuss the other two proposed methods, which turn out to be not experimentally feasible yet.

\subsection{Solar, atmospheric, and long-baseline neutrinos}

The current constraints on neutrino decay for a normal nondegenerate mass ordering are
\begin{subequations}
\begin{align}
\frac{\tau_3}{m_3}&> 2\times 10^{-10}\frac{{\rm s}}{{\rm eV}}\label{eq:atmdecay}\\
\frac{\tau_2}{m_2}&> 1.0\times 10^{-3}\frac{{\rm s}}{{\rm eV}} \ .\label{eq:solardecay}
\end{align}
\end{subequations}
The bound on $\tau_3$ at 90\% CL is obtained through an analysis of long-baseline neutrinos~\cite{Gago:2017zzy,Ascencio-Sosa:2018lbk} (comparable constraints can be put on invisible decays into sterile neutrinos, using atmospheric and long-baseline neutrino data \cite{GonzalezGarcia:2008ru}). The bound on $\tau_2$ at 99\% CL has been recently obtained with updated solar neutrino measurements \cite{Aharmim:2018fme}, which have improved the previous estimate \cite{Picoreti2015}.
It is important to notice that these bounds only apply to invisible neutrino decays, i.e., the decay products are assumed not to cause significant signals in the detectors \cite{GonzalezGarcia:2008ru,Picoreti2015}. Moreover, we stress again that these limits on the lifetimes assume a hierarchical ordering \cite{Beacom:2002cb,Berryman2014}. The constraints in Eqs.~\eqref{eq:atmdecay} and \eqref{eq:solardecay} require the off-diagonal couplings to obey
\begin{align}\label{eq:CouplingLimits}
\sqrt{g_{32}^2+g_{31}^2}< 9\times 10^{-2}\;\:\:\:\:\text{and}\;\:\:\:\:g_{21}< 3\times 10^{-4}\ .
\end{align}
These bounds on the couplings are valid both for Majorana neutrinos and, once multiplied by a factor of 2, also for Dirac neutrinos. In the Majorana case, additional strong bounds can be obtained because solar neutrinos would decay to antineutrinos showing up at experiments like SNO \cite{Aharmim:2004uf} and KamLAND \cite{Eguchi:2003gg}. Also Borexino~\cite{Bellini:2010gn} can put competitive bounds on the lifetime of solar neutrinos. However, the analysis in \cite{Bellini:2010gn} only yields bounds on the conversion of $\nu$ to $\overline{\nu}$ assuming either equal shapes for the solar $\nu$ and the converted $\overline{\nu}$ or a unknown shape for the $\overline{\nu}$ spectrum, while no dedicated analysis is carried out for neutrino decay. 

The analysis of KamLAND data gives bounds on the lifetime of $\nu_2$ Majorana neutrinos of \cite{Eguchi:2003gg}
\begin{subequations}
\begin{align}
\frac{\tau_2}{m_2}&> 6.7\times 10^{-2}\frac{{\rm s}}{{\rm eV}} \quad \rm (degenerate)
\\
\frac{\tau_2}{m_2}&> 1.1\times 10^{-3}\frac{{\rm s}}{{\rm eV}} \quad \rm (hierarchical)
\end{align}
\end{subequations}
at 90\% CL, which translate to bounds on the coupling
\begin{subequations}
\begin{align}
g_{21}&< 4\times 10^{-5}\quad \rm (degenerate)
\\
g_{21}&< 3\times 10^{-4} \quad \rm (hierarchical) \ .
\end{align}
\end{subequations}

Concerning $\nu_3$ decay, there is as yet no dedicated analysis in the context of majoron models \cite{Tanabashi:2018oca}, presumably because it was not known until recently that there is a small component of $\nu_3$ in electron neutrinos. To estimate the flux of $\nu_3$ from the Sun, we observe that this mass eigenstate is not affected by matter effects, so its component is simply $|U_{e3}|^2=0.02$ \cite{Capozzi:2018ubv}. In conclusion, approximately $2\%$ of the neutrinos coming from the Sun are $\nu_3$. The bounds on the lifetime of $\nu_3$ Majorana neutrinos are then
\begin{subequations}
\begin{align}
\frac{\tau_3}{m_3}&> 1.3\times 10^{-4}\frac{{\rm s}}{{\rm eV}} \quad \rm (degenerate)
\\
\frac{\tau_3}{m_3}&> 2.2\times 10^{-5}\frac{{\rm s}}{{\rm eV}} \quad \rm (hierarchical)
\end{align}
\end{subequations}
at 90\% CL, and the bounds on the coupling are finally
\begin{align}
&\sqrt{g_{32}^2+g_{31}^2}< 3\times 10^{-4}\quad \rm (degenerate)
\\
&\sqrt{g_{32}^2+g_{31}^2}< 2\times 10^{-3} \quad \rm (hierarchical).
\end{align}
These bounds apply to any model where neutrinos can decay to a light pseudoscalar.

\subsection{IceCube and supernova neutrinos}

Neutrino decays imply a distinct flavor composition of long-traveling astrophysical neutrinos because all neutrinos arrive in the lightest mass state. Observable decay effects require $\Gamma_i (m_i/E)\gtrsim D^{-1}$ and therefore \cite{Hannestad2005}
\begin{equation}\label{eq:gDecay}
g_{ij}\gtrsim5\times 10^{-8}\left(\frac{50{\rm ~meV}}{m_i}\right) \left(\frac{E}{{\rm 10~TeV}}\right)^{1/2} \left(\frac{{\rm 100~Mpc}}{D}\right)^{1/2} .
\end{equation}
As mentioned in Ref.~\cite{Dvali2016a}, the relatively weak constraints in Eq.~\eqref{eq:CouplingLimits} from atmospheric, long-baseline, and solar neutrinos therefore imply that a deviation from an equal neutrino flavor ratio $(\nu_e:\nu_\mu:\nu_\tau)=(1:1:1)$ could be measured at experiments such as IceCube.
While normal mass ordering would imply a dominance of $\nu_e$ due to $\nu_1=(0.68:0.11:0.21)$, inverted ordering would yield a dominance of $\nu_\mu$ and $\nu_\tau$ due to $\nu_3=(0.02:0.54:0.44)$~\citep{Capozzi:2018ubv}. An equal flavor ratio would not be allowed because only the intermediate eigenstate $\nu_2$ has an almost equal flavor content. 

An additional question is whether one can distinguish Majorana from Dirac neutrinos at IceCube. At low energies, the detector cannot distinguish neutrinos from antineutrinos. Nevertheless, for neutrinos with energies around $6.3 \, \rm PeV$ in the laboratory frame, the Glashow resonance makes IceCube more efficient for $\overline{\nu}_e$ detection~\cite{Nunokawa:2016pop}. Such a resonance is the s-channel of the process $\overline{\nu}_e+e^-\to W^- \to \overline{\nu}_\alpha + l^-_\alpha$ and allows the detector to distinguish neutrinos from antineutrinos. However, the main problem here is the unknown asymmetry of the neutrino and antineutrino fluxes. For analyzing IceCube data, it is customary to assume equal fluxes for ${\nu_\alpha}$ and ${\overline{\nu}_\alpha}$, both in energy and flavor \cite{Aartsen:2014gkd}. In this case, it would be impossible to distinguish Majorana from Dirac neutrinos. However, in general, these fluxes are expected to be different. Neutrinos are produced by the decay of charged pions via the decay chain \cite{Gaisser:2016uoy}
\begin{align}
\pi^{+}\to\, &\mu^{+}+\nu_\mu  \nonumber\\
&\downarrow \nonumber \\
 &e^+ +\nu_e+\overline{\nu}_\mu
\end{align}
and the corresponding charge-conjugate process. If there is an asymmetry in charged-pion production, there will be an asymmetry in neutrino fluxes. For example, the production mechanism $p+\gamma\to\Delta^+\to n+\pi^+$ involving proton collisions on photons of the environment ($p\gamma$ sources) produces no $\pi^-$ and thus no $\overline{\nu}_e$ \cite{Nunokawa:2016pop}. Note that the $\overline{\nu}_e$ from neutron beta decay can be neglected here due to the large neutron lifetime compared to its propagation timescale and the short $\pi^{+}$ and $\mu^{+}$ lifetimes.

For normal mass ordering, complete neutrino decay is currently disfavored by IceCube data at the $2\sigma$ level~\cite{Bustamante:2015waa}. Moreover, a single event detected in the energy range of the Glashow resonance could rule out complete decay for inverted mass ordering \cite{Bustamante:2016ciw}. Even though such bounds are stronger than the ones from solar neutrinos obtained in this paper, we stress that solar neutrinos already allow one to distinguish Majorana from Dirac neutrinos, making the results previously discussed complementary to IceCube results. In the far future, a better understanding of neutrino production in astrophysical sources would give us tools to further distinguish the Majorana from the Dirac scenario.

The same might be true for supernova neutrinos, which are expected to decay into the lightest mass eigenstate while traveling to Earth. While this decay scenario requires modified analyses of original supernova neutrino spectra \citep{Dvali2016a}, it could also be probed through the future detection of the supernova relic neutrino flux, i.e., the redshifted neutrino background from all past supernovae. In Ref.~\cite{Fogli2004} it was argued that a complete decay scenario can potentially enhance the supernova relic neutrino background density up to the current experimental detection bound, so that its measurement might be feasible with near-future experiments.

\section{Conclusions}\label{sec:conclusions}

In this paper, we have explored the possibility of distinguishing Majorana from Dirac neutrinos by their gravi-majoron decays, assuming a gravitational origin of neutrino masses as proposed in Ref.~\citep{Dvali2016a}. Interactions between neutrinos and conventional Nambu-Goldston bosons are strongly constrained by cosmology, astrophysics, and laboratory experiments. In contrast, the gravi-majorons we consider are not ruled out because they only arise in the very late Universe and effectively decouple from high-energy processes. The detection of an unexpectedly large absolute neutrino mass scale in beta-decay experiments could provide a hint toward the gravitational mass mechanism because it evades all cosmological mass bounds. Such a large absolute neutrino mass scale would imply a degenerate mass hierarchy, allowing us to distinguish Majorana from Dirac neutrinos through their decay. In the Majorana case, we put strong quantitative bounds on the decay of the heaviest neutrino, ${\tau_3}/{m_3}> 2.2\times 10^{-5}{{\rm s}}/{{\rm eV}}$ at 90\% CL, using data from experiments searching for antineutrino appearance from the Sun. We also qualitatively explored the possibility of distinguishing Dirac from Majorana neutrinos with future IceCube data; however, such a possibility requires a better understanding of neutrino production in astrophysical sources. The future will bring more data in neutrino astronomy, and perhaps even a supernova, to further study such an exotic, yet fascinating, scenario.

\section*{ACKNOWLEDGMENTS}

We thank Gia Dvali and Aaron Vincent for discussions and acknowledge partial support by the Deutsche Forschungsgemeinschaft through SFB 1258 (Collaborative Research Center ``Neutrinos, Dark Matter,
Messengers''), as well as the European Union through Grant
No.\ H2020-MSCA-ITN-2015/674896 (Innovative Training Network
``Elusives''). Research at Perimeter Institute is supported by the Government of Canada through the Department of Innovation, Science and Economic Development and by the Province of Ontario through the Ministry of Research, Innovation and Science.

\appendix

\section{High-energy softening of the gravitational vertex}\label{app:softening}

In this appendix, we explain and constrain the high-energy softening of the nonperturbative gravitational vertex that gives rise to the coupling in Eq.~\eqref{eq:nudecay}. Some of the constraints have already been discussed in Refs.~\citep{Dvali2016a,Dvali2016b}.

\subsection{Cosmological constraints}

In the early Universe before photon decoupling, i.e., for $T\gtrsim 0.3~{\rm eV}$, neutrino self-interactions $\lambda_{ij}=\{g_{ij},y_{ij}\}$ are strongly constrained by data from the cosmic microwave background, $\lambda_{ij}\lesssim 10^{-7}$ \cite{Hannestad2005,Archidiacono2013}. Since we only compare relative temperatures here, we can neglect the ratio of $(4/11)^{1/3}$ between the neutrino and photon temperatures. Moreover, in the even earlier Universe before neutrino freeze-out, i.e., for $T\gtrsim 1~{\rm MeV}$, the existence of new degrees of freedom (e.g., sterile neutrinos) is excluded if they had been in equilibrium with photons or electrons~\cite{Aghanim:2018eyx}. Thus, the interaction rate $\Gamma = \langle\sigma_\nu v_\nu\rangle n_\nu \sim \lambda_{ij}^4 T$ needs to be smaller than the Hubble expansion rate $H(T) \sim T^2/M_P$, which yields the constraint $\lambda_{ij} < (T/M_P)^{1/4} \sim 10^{-8}$ for $T={\rm MeV}$. Note that new neutrino physics at lower temperatures becomes irrelevant for this constraint, since energy is conserved in the relic neutrino sector after neutrinos freeze out.

The mentioned cosmological constraints do not rule out the 't-Hooft-like gravitational four-neutrino vertex~\citep{Dvali2016a,Dvali2016b} that gives rise to the interactions $\lambda_{ij}$ in Eq.~\eqref{eq:nudecay} after the phase-transition, because this effective low-energy vertex is strongly temperature-suppressed before the neutrino phase-transition, $\lambda_{ij}(T)\propto (\Lambda_G/T)^n$.\footnote{Note that the analogous gravitational four-fermion vertices for charged fermions are additionally suppressed by the masses of these fermions, which renders them cosmologically irrelevant~\citep{Dvali2016a}.} Here, $n$ is an unknown power-law exponent that has to be fixed by phenomenological requirements.

For comparison, the nonperturbatively generated QCD axion mass is suppressed by $m_a\propto (\Lambda_{\rm QCD}/T)^{11}$ for $T\gg m_q$, where $\Lambda_{\rm QCD}$ is the QCD scale and $m_q$ are the quark masses \cite{Sikivie:2006ni}. However, we note that the analogy between the QCD and gravitational scenarios does not go too far because the perturbative QCD coupling becomes strong at $T\sim\Lambda_{\rm QCD}$, while the perturbative gravitational coupling is expected to be weak and cosmologically decoupled at the phase-transition temperature $T_G\sim\Lambda_G$~\cite{Dvali2016a,Dvali2016b}.
While quarks and pions couple via the slowly running perturbative QCD coupling, our gravitational vertex is purely nonperturbative and thus expected to decrease much faster than in the perturbative QCD case. Thus, our analogy is only based on possible similarities between the nonperturbative and anomaly structures of these two theories \citep{Dvali2016a}.

The mentioned cosmological constraints require $\lambda_{ij}(T_0)\lesssim 10^{-4}$ for a minimal suppression of $n=1$ and a phase-transition temperature of $T_G\gtrsim T_0$, where $T_0 \sim 0.2$~meV is the cosmic temperature today. Already for $n\geq 2$, the zero-temperature couplings can be as large as $\lambda_{ij}(T_0)\sim 1$. We note in this context that a simplified cosmological analysis of the neutrino mass model demonstrates that the phase-transition is favored to happen at very low redshifts, $z\sim 0$ \cite{Lorenz:2018fzb}.

\subsection{Stellar and laboratory constraints}

The gravi-majoron production in high-energy astrophysical environments with $E\gg m_\nu$ is suppressed by $(\Lambda_G/E)^{n}$, where $n=1$ is already sufficient to satisfy the astrophysical constraints on $g_{\phi\gamma\gamma}$ mentioned in Sec.\ \ref{sec:astrobounds}. The gravitational $\nu\overline{\nu}\gamma$, $\nu\overline{\nu}\phi$, $\nu\overline{\nu}\nu\overline{\nu}$, $e\overline{e}\phi$, and $e\overline{e}\nu\overline{\nu}$ couplings are also strongly suppressed in stellar processes. However, the constraints on direct couplings to electrons are stronger than in the two-photon case, due to the absence of the electron-loop suppression. The most stringent bounds on majoron-like models usually come from Compton or bremsstrahlung processes in red giant cores, $g_{e\overline{e}\phi}\lesssim 10^{-9}~{\rm GeV^{-1}}$ \citep{Raffelt:1994ry}, which equally apply to $g_{e\overline{e}\nu\overline{\nu}}$. In our case, these astrophysical bounds cannot be trivially translated into bounds on $n$, since astrophysical processes typically involve different momentum transfers. In particular, these processes becomes less $n$-suppressed but more phase-space suppressed at low momentum transfer \cite{Raffelt:1996wa}. We defer a detailed analysis of such constraints to later studies and only observe here that the constraints should be satisfied for a relatively low suppression factor.

The most stringent laboratory bounds on fifth forces mediated by the scalar analogs of gravi-majorons \cite{Dvali2016b} are $|\alpha|\lesssim 10^6$ $(10^{-2})$ for $\Lambda_G\sim 0.1~$eV (1~meV) \cite{Chen:2014oda}, where $|\alpha|$ is the strength of the Yukawa-like correction to the Newtonian gravitational constant (i.e., normalized to gravity). These bounds on fifth forces between hadrons translate into constraints on the power-law exponent of the large-mass decoupling, $n\gtrsim 1.6$ $(1.7)$, due to $|\alpha|\sim (\Lambda_G/m_p)^{2n}(M_P/m_p)^2$ at distances $r\sim \Lambda_G^{-1}$ \citep{Dvali2016b}. Here, we used $(\Lambda_G/m_p)\sim 10^{-10}$ $(10^{-12})$ and denoted $M_P$ and $m_p$ as the Planck and proton masses, respectively.
The high-energy suppression of the predicted fifth force between electrons is much weaker due to $(\Lambda_G/m_e)\sim 2\times 10^{-7}$ $(2\times 10^{-9})$, but the experimental bounds on such a force are weaker as well, $|\alpha|\lesssim 10^{22}$ ($10^{18}$) \citep{Wise:2018rnb}, translating to $n\gtrsim 1.2$ ($1.2$). Thus, all of the laboratory constraints are compatible with the model predictions for $n\gtrsim 2$ \citep{Dvali2016b}.

\section{Matrix element for neutrino decay}\label{appendix}
In this appendix, we calculate the matrix element for the decay process $\nu_i\to
\overset{\barparen}{\nu}_j \, +\, \phi$ of a Dirac or a Majorana neutrino with fixed initial spin and derivative coupling to a gravi-majoron\footnote{Notice there is a special case in which the relative CP phase of the decaying neutrino and the daughter neutrino is either $-1$ or $+1$. If this scenario is realized, Majorana neutrino can only have either pseudoscalar or pseudovector coupling, not both.}
\begin{align}\label{eq:nudecayappend}
 \mathcal{L}_{\rm int} =  \partial_\mu \phi \sum_{ij} \frac{g_{ij}}{m_i+m_j} \overline{\nu}_i \gamma^\mu\gamma_5 \nu_j + \mathrm{h.c.} 
\end{align}
For the Dirac case, fixing the initial spin is equivalent to choosing an active {\it or\/} a sterile initial state, whereas for a Majorana neutrino it simply means choosing if it is a neutrino (left-handed) or an antineutrino (right-handed).
Using the Feynman rules of Ref.~\cite{Gluza:1991wj}, we find
\begin{subequations}\label{eq:matrixelement}
\begin{align}
  {\cal M}^{\rm D} &=
  g_{\phi,ij}\overline u_{\nu_j}\gamma^\mu\gamma^5u_{\nu_i}J^{\phi}_\mu\, &&{\rm (Dirac)} \\
     {\cal M}^{\rm M} &=  g_{\phi,ij}\overline u_{\nu_j}\gamma^\mu\gamma^5u_{\nu_i }J^{\phi}_\mu \times 2\,   &&{\rm (Majorana),}
\end{align}
\end{subequations}
where we have neglected global phases and $
J^{\phi}_\mu~=~k_\mu$
is the gravi-majoron current.
Note that the matrix element for the Majorana case  is twice as large as for the Dirac case because the Hermitian conjugate in the Lagrangian also contributes to the amplitude, so that the rate will be four times larger for a decaying Majorana neutrino. This can be compared, e.g., to radiative decays induced by standard model interactions, for which the one of Majorana neutrinos is two times larger than the decay width of Dirac neutrinos \cite{Gorbunov:2007ak}. The global constant can be reabsorbed in the coupling definition.

Let us consider the case of a decaying Majorana neutrino. Squaring the amplitude one finds
\begin{align}\label{squaredamplitude}
  |{\cal M}^{\rm M} |^2 &= \ 
  \frac{g_{ij}^2}{(m_i+m_j)^2} \mathrm{Tr}\[(\slashed{p}_i+m_i)(1+h_i\gamma^5 \slashed{S}_i)\gamma^\mu \gamma^5\right.
  \nonumber \\ &\left.
  \times(\slashed{p}_j+m_j)(1+h_j\gamma^5 \slashed{S}_j)\gamma^\nu \gamma^5\]k_\mu k_\nu \ ,
\end{align}
where $p_i$, $p_j$, and $k$ are the four-momenta of $\nu_i$, $\nu_j$, and $\phi$ respectively and we used~\cite{Long:2014zva}
\begin{equation}
u_\nu \overline u_\nu=\frac 1 2 (\slashed
p+m_\nu)\left(1+h\gamma^5\slashed S \right)
\end{equation}
with the spin vector
\begin{equation}\label{spin}
S=\left(\frac{|{\bf p}|}{m_\nu}\, ,\, \frac{E_\nu}{m_\nu}\hat{\bf
p}\right)\,.
\end{equation}
The squared amplitude in Eq.~\eqref{squaredamplitude} has 16 terms, 8 of which contain an odd number of $\gamma^\mu$ and 4 of which have a $\gamma^5$ so that they do not contribute for symmetry reasons. The remaining terms give
\begin{align}\label{squaredamplitudefinal}
  |{\cal M}^{\rm M} |^2 = \,
  4 \, g_{ij}^2
 \[(p_i \cdot p_j-m_i m_j)(1+h_i h_j S_i\cdot S_j)\right.
   \nonumber \\ \left.
{} -h_i h_j(p_i\cdot S_j)(p_j\cdot S_i)\] \ .
\end{align}
Analogously, one can compute the decay through a pseudoscalar coupling
\begin{align}
 \mathcal{L}_{\rm int} = i\phi \sum_{ij} g_{ij} \overline{\nu}_i \gamma_5 \nu_j+\rm h.c. 
\end{align}
 and find explicitly that it has the same squared amplitude as in Ref.~\cite{Kim:1990km}. Notice that there are two factors to be taken into account. The coupling in our interaction Lagrangian is twice as large as in Ref.~\cite{Kim:1990km}; furthermore, we use a different spinor normalization, so an additional factor $1/(4 \, m_i m_j)$ has to be included. Our results agree with Ref.~\cite{Kim:1990km} but differ from the ones reported in Ref.~\cite{Beacom:2002cb}, where the couplings are defined as in our Lagrangian but the rate is quoted directly from Ref.~\cite{Kim:1990km}; the results in Ref.~\cite{Beacom:2002cb} would be correct if they considered exclusively Dirac neutrinos. In the following we will consider only the pseudoscalar coupling, as the derivative coupling process is equivalent to the latter.

After substituting Eq.~\eqref{spin} in~\eqref{squaredamplitudefinal}, one can evaluate the squared amplitude in the $\nu_i$ rest frame (where the spin vector $S=(0,\bold{S})$, so that $S\cdot p=0$), showing that the emission of a certain helicity is not isotropic,
\begin{align}\label{squaredamplitudefinal}
  |{\cal M}^{\rm M} |^2 &= \,
  4 \, g_{ij}^2  \[(m_i E_j-m_i m_j)\(1-h_i h_j E_j\frac{\bold{p}_j}{|\bold{p}_j|}\cdot \bold{S}_i \)\right.
   \nonumber \\ &\left.
 - h_i h_j\(m_i\frac{|\bold{p}_j|}{m_j}\)(\bold{p}_j\cdot \bold{S}_i)\]  \ .
\end{align}
The different angular distributions in the rest frame translate
into different energy spectra in the laboratory frame for different
helicities~\cite{Balantekin:2018ukw}. We can evaluate the squared amplitudes in the laboratory frame, in which $E_i\gg m_i$, finding for the helicity conserving decay
\begin{align}
  |{\cal M}^{\rm M}_{\nu_i\to \nu_j+\phi}|^2 &=4 \, g_{ij}^2 m_i m_j\(A-2\)
\end{align}
and for the helicity flipping decay
\begin{align}
  |{\cal M}^{\rm M}_{\nu_i\to \overline{\nu}_j+\phi}|^2 &=4 \, g_{ij}^2 m_i m_j\(\frac{m_i^2+m_j^2}{m_i m_j}-A\)\ ,
\end{align}
where
\begin{align}
A\equiv \( \frac{m_i E_j}{m_j E_i}+\frac{m_j E_i}{m_i  E_j }     \) \ .
\end{align}
The decay rate for Dirac neutrinos to an active or a sterile neutrino neutrino are found by dividing these expression by 4 and interpreting $\overline{\nu}_j$ as a sterile state $N_j$. These are the expressions that are relevant for the decay rate discussed in the main text.

The vector-like coupling case is obtained following the same steps. Starting from the Lagrangian
\begin{align}
 \mathcal{L}_{\rm int} =  \partial_\mu \phi \sum_{i\neq j} \frac{g_{ij}}{m_i-m_j} \overline{\nu}_i \gamma^\mu \nu_j + \mathrm{h.c.} 
\end{align}
one finds for the helicity conserving decay
\begin{align}
  |{\cal M}^{\rm M}_{\nu_i\to \nu_j+\phi}|^2 &=4 \, g_{ij}^2 m_i m_j\(A+2\)
\end{align}
and for the helicity flipping decay
\begin{align}
  |{\cal M}^{\rm M}_{\nu_i\to \overline{\nu}_j+\phi}|^2 &=4 \, g_{ij}^2 m_i m_j\(\frac{m_i^2+m_j^2}{m_i m_j}-A\)\ .
\end{align}
The decay rate for Dirac neutrinos are again found by dividing these expression by 4 and interpreting $\overline{\nu}_j$ as a sterile state $N_j$. 

\bibliographystyle{bibi}
\bibliography{bibliography}

\end{document}